\documentstyle[12pt,epsfig]{article}

\begin{document}
\begin {center}
{\bf {\Large
Nuclear transparency of the charged hadron produced in the inclusive
electronuclear reaction
} }
\end {center}
%\smallskip
%\medskip
\begin {center}
Swapan Das \footnote {email: swapand@barc.gov.in} \\
{\it Nuclear Physics Division,
Bhabha Atomic Research Centre,  \\
Trombay, Mumbai 400085, India \\
Homi Bhabha National Institute, Anushakti Nagar,
Mumbai 400094, India }
\end {center}
%\medskip

\begin {abstract}
The nuclear transparency of the charged hadron produced in the inclusive
$(e,e^\prime)$ reaction on the nucleus has been calculated using Glauber
model for the nuclear reaction.
The
color transparency (CT) of the produced hadron and the short-range
correlation (SRC) of the nucleons in the nucleus have been incorporated in
the Glauber model to investigate their effects on the nuclear transparency
of the hadron. The calculated nuclear transparencies for the proton and pion
are compared with the data.
\end {abstract}

%Keywords:
%electronuclear reaction, color transparency, short-range correlation

%PACS number(s): 13.60.Le, 25.20.-x

%\section{Introduction}

The hadron-nucleus cross section is less than that in the plane wave impulse
approximation (PWIA) because the initial and(or) final state interactions of
the hadron with the nucleus are neglected in PWIA. The difference in the
cross sections can be characterized by the nuclear transparency $T_A$,
defined \cite{Howell} as
\begin{equation}
T_A= \frac{\sigma_{hA}}{\sigma_{hA (PWIA)}},
\label{TAxn}
\end{equation}
where $\sigma_{hA}$ represents the hadron-nucleus cross section.

The transverse size $d_\perp$ of the hadron produced in the nucleus due to
the space-like high momentum transfer $Q^2$ is reduced as $d_\perp \sim 1/Q$
\cite{Howell, Farrar}. The reduced (in size) hadron is referred as point like
configuration (PLC) \cite{Howell}.
According
to Quantum Chromodynamics, a color neutral PLC has reduced interaction with
the nucleon in the nucleus because the sum of its gluon emission amplitudes
cancel \cite{Howell, DuttaH}.
The PLC expands to the size of physical hadron, as it moves up to a length
($\sim 1$ fm) called hadron formation length $l_h$ \cite{Howell, Larionov}:
\begin{equation}
l_h= \frac{2k_h}{\Delta M^2},
\label{Lh}
\end{equation}
where $k_h$ is the momentum of hadron in the laboratory frame. $\Delta M^2$
is related to the mass difference between the hadronic states originating
due to the (anti)quarks fluctuation in PLC.
The
interaction of PLC with the nucleon in the nucleus increases, as its size 
increases during its passage $l_h$ through the nucleus.
The
decrease in the hadron-nucleon cross section in the nucleus, as explained
by Glauber model \cite{Glauber}, leads to the increase in the hadron-nucleus
cross section $\sigma_{hA}$. Therefore, the transparency $T_A$ in
Eq.~(\ref{TAxn}) of the hadron raises.
The
enhancement in $T_A$ due to the above phenomenon is referred as color
transparency (CT) of the hadron. The physics of CT for hadrons have been
discussed elaborately in Refs.~\cite{DuttaH, Frankfurt}.

The experiments to search the CT of proton ($p$CT) in the A$(p,pp)$ reactions
done at Brookhaven National Laboratory (BNL) \cite{Carroll} could not confirm
the $p$CT. In fact, the experimental results are not understood \cite{Lee}.
The
$p$CT is not also seen in the A$(e,e^\prime p)$ experiment done at Standford
Linear Accelerator Center (SLAC) \cite{Neill} and Jefferson Laboratory (JLab)
\cite{DuttaG} for $0.64 \le Q^2 \le 8.1$ GeV$^2$.
This
experiment done for $8 \le Q^2 \mbox{(GeV$^2$)} \le 14.2$ \cite{Bhetuwal} at
the upgraded JLab facility agrees with the previous observation
\cite{Neill, DuttaG}.
Therefore,
it appears the PLC required for CT is unlikely to form for three quarks
$(qqq)$ system, such as proton.

Since the meson is a bound state of two quarks (i.e., quark-antiquark) the
PLC formation of it can be more probable than that of the baryon, a three
quarks $(qqq)$ system.
The
color transparency is unambiguously reported from Fermi National Accelerator
Laboratory (FNAL) \cite{Aitala} in the experiment of the nuclear diffractive
dissociation of pion (of 500 GeV/$c$) to dijets.
The
color transparency is also illustrated in the $\pi^-$ meson photoproduction
\cite{DuttaE} and $\rho^0$ meson electroproduction (from nuclei) experiments
\cite{Airapetian}.
Several
authors studied the $\rho$-meson color transparency in the energy region
available at JLab \cite{Howell, Kopeliovich}.

The nuclear transparency of the $\pi^+$ meson produced in the
A$(e,e^\prime)$ process was measured at JLab for the photon virtuality
$Q^2=1.1-4.7$ GeV$^2$ \cite{Clasie}. The data have been understood by the
pionic color transparency ($\pi$CT) \cite{Kaskulov}. Larson et al.,
\cite{Larson} described the momentum dependence of $\pi$CT in the above
reaction.
Cosyn
et al., \cite{cosyn} studied the effects of $\pi$CT and nucleon short-range
correlation in the pion photo- and electro- production from nuclei.
Larionov
et al., \cite{Larionov} estimated the $\pi$CT in the $(\pi^-,l^+l^-)$
reaction on nuclei for $p_\pi = 5-20$ GeV/$c$, which can be measured at the
forth-coming facilities in Japan Proton Accelerator Research Complex
(J-PARC) \cite{Kumano}. This reaction provides the informations complementary
to those obtained from the A$(\gamma^*,\pi)$ reaction.
Miller
and Strikman \cite{MillerS} illustrated large CT in the pionic knockout of
the proton off nuclei at the energy 200 GeV available at CERN COMPASS
experiment.

The enhancement in $T_A$ due to $\sigma_{hA}$ in Eq.~(\ref{TAxn}) can also
occur because of the short-range correlation (SRC) of nucleon in the nucleus.
The SRC arises because of the repulsive (short-range) interaction between
the nucleons bound in the nucleus.
This
interaction keeps the bound nucleons apart ($\sim$ 1 fm), which is called
nuclear granularity \cite{Lee}. Therefore, the SRC prevents the shadowing
of the hadron-nucleon interaction due to the surrounding nucleons present
in the nucleus. This occurrence, as elucidated by Glauber model
\cite{Glauber}, leads to the enhancement in $\sigma_{hA}$.
The SRC is widely used to investigate various aspects in the nuclear physics
\cite{MillerJ}.

%\section{Formalism}

The hadron $h$ is produced  in the inclusive A$(e,e^\prime)$X reaction
because of the interaction of the virtual photon $\gamma^*$ (emitted at the
$ee^\prime$ vertex) with the nucleus A. In this reaction, the nucleus in the
final state denoted by X is unspecified.
The
scattering amplitude for the $\gamma^* \mbox{A} \to h \mbox{X}$ transition,
according to Glauber model \cite{Howell}, can be written as
\begin{equation}
F_{X0} [({\bf{q-k}}_h)_\perp]
=\frac{iq}{2\pi}
\int d{\bf b} e^{i({\bf{q-k}}_h)_\perp \cdot {\bf {b}}}
\Gamma^{\gamma^*h}_{X0} ({\bf b}),
\label{FX01}
\end{equation}
where ${\bf q}$ and ${\bf k}_h$ are the momenta of $\gamma^*$ and $h$
respectively. $\Gamma^{\gamma^*h}_{X0} ({\bf b})$ describes the matrix
element for the transition of the nucleus from its initial to final states,
i.e.,
\begin{equation}
\Gamma^{\gamma^*h}_{X0} ({\bf b})
= <X|\Gamma^{\gamma^*h}_A ( {\bf b}, {\bf r}_1, ..., {\bf r}_A )|0>,
\label{GX01}
\end{equation}
where $|0>$ denotes the ground state of the target nucleus and $|X>$
represents the unspecified nuclear state in the exit channel. The nuclear
profile operator
$\Gamma^{\gamma^*h}_A ( {\bf b}, {\bf r}_1, ..., {\bf r}_A )$
\cite{Howell, Hufner} is given by
\begin{equation}
\Gamma^{\gamma^*h}_A ( {\bf b}, {\bf r}_1, ..., {\bf r}_A )
= \sum_i \Gamma^{\gamma^*h} ( {\bf b-b}_i )
e^{i ({\bf q-k}_h)_\parallel z_i }
\Pi^{A-1}_{j\not{=}i} [1 -\Gamma^{hN}({\bf b-b}_j) \theta(z_j-z_i)].
\label{GA01}
\end{equation}
The summation $i$ is taken over the number of nucleons in the nucleus
participated for the hadron production, e.g., the protons in the nucleus
take part to produced charged hadron in the reaction.

$\Gamma^{\gamma^*h} ( {\bf \tilde {b}} )$ is the two-body profile function
for the hadron produced from the nucleon, i.e., $\gamma^* N \to hN$ process.
It is related to the reaction amplitude
$f_{\gamma^*h} ({\bf \tilde {q}}_\perp)$ \cite{Howell} as
\begin{equation}
\Gamma^{\gamma^*h} ({\bf \tilde {b}})
= \frac{1}{i2\pi q}
\int d{\bf \tilde {q}}_\perp e^{-i{ \bf \tilde {q}}_\perp \cdot
{\bf \tilde {b}} } f_{\gamma^*h} ({\bf \tilde {q}}_\perp).
\label{GN01}
\end{equation}
The two-body profile function $\Gamma^{hN}({\bf \tilde {b}})$ is connected
to $hN$ (hadron-nucleon) elastic scattering amplitude
$f_{hN} ({\bf \tilde {q}}_\perp)$ \cite{Howell, Glauber} as
\begin{equation}
f_{hN} ({\bf \tilde {q}^\prime}_\perp)
= \frac{ik_h}{2\pi}
\int d{\bf \tilde {b}^\prime} e^{ i{ \bf \tilde {q}^\prime}_\perp \cdot
{\bf \tilde {b}^\prime} } \Gamma^{hN} ({\bf \tilde {b}^\prime}).
\label{GN02}
\end{equation}

The nuclear states, assuming the independent particle model \cite{Bauer},
can be written in terms of the single particle state $\Phi$ as
$|0>= \Pi^A_{l=1} |\Phi_0 ({\bf r}_l)>$ and
$|X>= |\Phi_X({\bf r}_m)> \Pi^{A-1}_{n {\not =} m} |\Phi_0 ({\bf r}_n)>$.
Using those, $\Gamma^{\gamma^*h}_{X0} ({\bf b})$ in Eq.~(\ref{GX01}) can be
written as
\begin{equation}
\Gamma^{\gamma^*h}_{X0} ({\bf b})=
\sum_i \int d{\bf r}_i \Phi^*_X({\bf r}_i)  \Gamma^{\gamma^*h} ({\bf b-b}_i)
e^{i ({\bf q-k}_h)_\parallel z_i} \Phi_0({\bf r}_i)  D({\bf b},z_i),
\label{Fn02}
\end{equation}
where $D({\bf b},z_i)$ is given by
\begin{eqnarray}
D({\bf b},z_i)
&=& 
\Pi^{A-1}_{j {\not =} i} \int d{\bf r}_j  \Phi^*_0 ({\bf r}_j)
[ 1 - \Gamma^{hN} ({\bf b-b}_j) \theta (z_j-z_i) ] \Phi_0 ({\bf r}_j)
\nonumber  \\
&=&
\left [ 1- \frac{1}{A} \int d{\bf b}_j \Gamma^{hN} ({\bf b-b}_j)
\int dz_j \theta (z_j-z_i) \varrho ({\bf r}_j) \right ]^{A-1}.
\label{Db01}
\end{eqnarray}
In this equation, $\varrho ({\bf r}_j)$ is the matter density distribution
of the nucleus, i.e., $\varrho ({\bf r}_j) = A |\Phi_0({\bf r}_j)|^2$.
$\varrho ({\bf b}_j, z_j)$ can be replaced by $\varrho ({\bf b}, z_j)$,
since $\Gamma^{hN} ({\bf b-b}_j)$ varies much rapidly than
$\varrho ({\bf b}_j, z_j)$ \cite{Howell}.
Using Eq.~(\ref{GN02}) and
${\cal L}t_{n \to \infty} ( 1 + \frac{x}{n} )^n =e^x$, the above equation
can be simplified to
\begin{equation}
D({\bf b},z_i) \simeq
e^{ -\frac{1}{2} \sigma^{hN}_t [1- i\alpha_{hN}] T({\bf b},z_i)},
\label{Db02}
\end{equation}
where $\alpha_{hN}$ denotes the ratio of the real to imaginary part of the
hadron-nucleon scattering amplitude $f_{hN}(0)$, and
$\sigma^{hN}_t$ = $\frac{4\pi}{k_h} Im[f_{hN}(0)]$ is the hadron-nucleon
total cross section. $T({\bf b},z_i)$ is the partial thickness function of
the nucleus, i.e.,
\begin{equation}
T({\bf b},z_i) = \int^\infty_{z_i} dz_j\varrho ({\bf b}, z_j).
\label{Tbz}
\end{equation}

Using Eq.~(\ref{Fn02}), $F_{X0} [({\bf{q-k}}_h)_\perp]$ in Eq.~(\ref{FX01})
can be expressed as
\begin{eqnarray}
F_{X0} 
&=&
\frac{iq}{2\pi} \int d{\bf b} e^{i({\bf{q-k}}_h)_\perp \cdot {\bf {b}}}
\sum_i \int d{\bf r}_i \Phi^*_X({\bf r}_i)  \Gamma^{\gamma^*h} ({\bf b-b}_i)
e^{i ({\bf q-k}_h)_\parallel z_i} \Phi_0({\bf r}_i)  D({\bf b},z_i),
\nonumber  \\
&=&
\sum_i \int d{\bf r}_i \Phi^*_X({\bf r}_i) f^{(i)}_{hN}([{\bf{q-k}}_h]_\perp)
e^{i ({\bf q-k}_h) \cdot {\bf r}_i} \Phi_0({\bf r}_i) D({\bf r}_i).
\label{Fn03}
\end{eqnarray}
$f^{(i)}_{hN}$, defined in Eq.~(\ref{GN02}), can be considered identically
equal for all nucleons.

%\section{Result and Discussions}

The nucleus in the final state $|X>$ differs from its initial state $|0>$
(i.e., ground state) for the charged hadron production, i.e.,
$\Phi_X {\not =} \Phi_0$ and $F_{00}=0$.
To
calculate the cross section, $|F_{X0}|^2$ is to multiply by the phase-space
of the reaction and that is to divide by the incident flux.
Since
the final state $|X>$ of the nucleus is not detected in the inclusive
reaction, the summation over all states $X$ has to carry out.
In
the multi-GeV region, the phase space of the reaction can be considered
independent of the state $X$, and therefore, the nuclear transparency $T_A$
can be written \cite{Howell} as
\begin{equation}
T_A =\frac{ \sum_{X \not {=} 0} |F_{X0}|^2 }
          { \sum_{X \not {=} 0} |F_{X0}|^2_{PWIA} }.
\label{TApF}
\end{equation}

The hadron-nucleon cross section $\sigma^{hN}_t$ in the free-space is used
in Eq.~(\ref{Db02}) to evaluate $T_A$ in Glauber model. To look for the
color transparency (CT), $\sigma^{hN}_t$ (according to quantum diffusion
model \cite{Farrar, Larson}) has to replace by $\sigma^{hN}_{t,CT}$:
\begin{equation}
\sigma^{hN}_{t,CT} (Q^2, l_z) =\sigma^{hN}_t
\left [
\left \{
\frac{l_z}{l_h} +\frac{n_q^2<k^2_t>}{Q^2} \left ( 1 -\frac{l_z}{l_h} \right )
\right \} \theta(l_h-l_z)
+ \theta(l_z-l_h) \right ],
\label{XCT}
\end{equation}
where $Q^2$ is the space-like four-momentum transfer, i.e., photon virtuality.
$n_q$ denotes the number of valence quak-(anti)quark present in the hadron,
e.g., $n_q=2(3)$ for pion (proton) \cite{Farrar}. $k_t$ illustrates the
transverse momentum of the (anti)quark: $<k_t^2>^{1/2} =0.35$ GeV/$c$.
$l_z$
is the path length traversed by the hadron after its production. The hadron
formation length $l_h (\propto \frac{1}{\Delta M^2})$ is already defined in
Eq.~(\ref{Lh}).

The short-range correlation (SRC) can be incorporated by replacing the
nuclear density distribution $\varrho$ in Eq.~(\ref{Tbz}) by
\begin{equation}
\varrho ({\bf b}, z_j) \to \varrho ({\bf b}, z_j) C(|z_j-z_i|),
\label{dcr}
\end{equation}
where $C(u)$ represents the correlation function \cite{Lee}. Using the
nuclear matter estimate, it can be written as
\begin{equation}
C(u) = \left [ 1-\frac{h(u)^2}{4} \right ]^{1/2} [1+f(u)],
\label{Crfn}
\end{equation}
with $h(u)=3\frac{j_1(k_Fu)}{k_Fu}$ and
$f(u)=-e^{-\alpha u^2} (1-\beta u^2)$.
The Fermi momentum $k_F$ is chosen equal to 1.36 fm$^{-1}$. $C(u)$ with the
parameters $\alpha =1.1$ fm$^{-2}$ and $\beta =0.68$ fm$^{-2}$ agrees well 
that derived from the many-body calculations \cite{Lee}.

The nuclear transparency $T_A$ of the charged hadron, i.e., proton and
$\pi^+$ meson, produced in the inclusive electronuclear reaction has been
calculated using Glauber model (GM), where the measured nuclear density
distribution $\varrho (r)$ \cite{Jager}, and hadron-nucleon cross section
$\sigma_t^{hN}$ \cite{Zyla} are used.
As
shown later, the calculated results due to GM (presented by the dashed
curves) underestimate the measured $T_A$ for both proton and pion.
Therefore, GM has been modified by taking account of CT and SRC.
Since
the CT is energy dependent, the calculated $T_A(\pi^+)$ increases with
$Q^2$ due to the inclusion of CT in GM. Unlike CT, the SRC is independent
of energy. Therefore, the calculated $T_A(\pi^+)$ due to SRC added in GM
does not illustrate the $Q^2$ dependence.
The
dot-dot-dashed and dot-dashed curves arise because of the inclusion of CT in
GM for $\Delta M^2$, defined in Eq.~(\ref{Lh}), taken equal to 0.7 and 1.4
GeV$^2$ respectively. The calculated $T_A$ due to SRC incorporated in GM are
presented by the solid curves.

\begin{figure}[h]
%\begin{figure}
\begin{center}
\centerline {\vbox {
%\psdraft
\psfig{figure=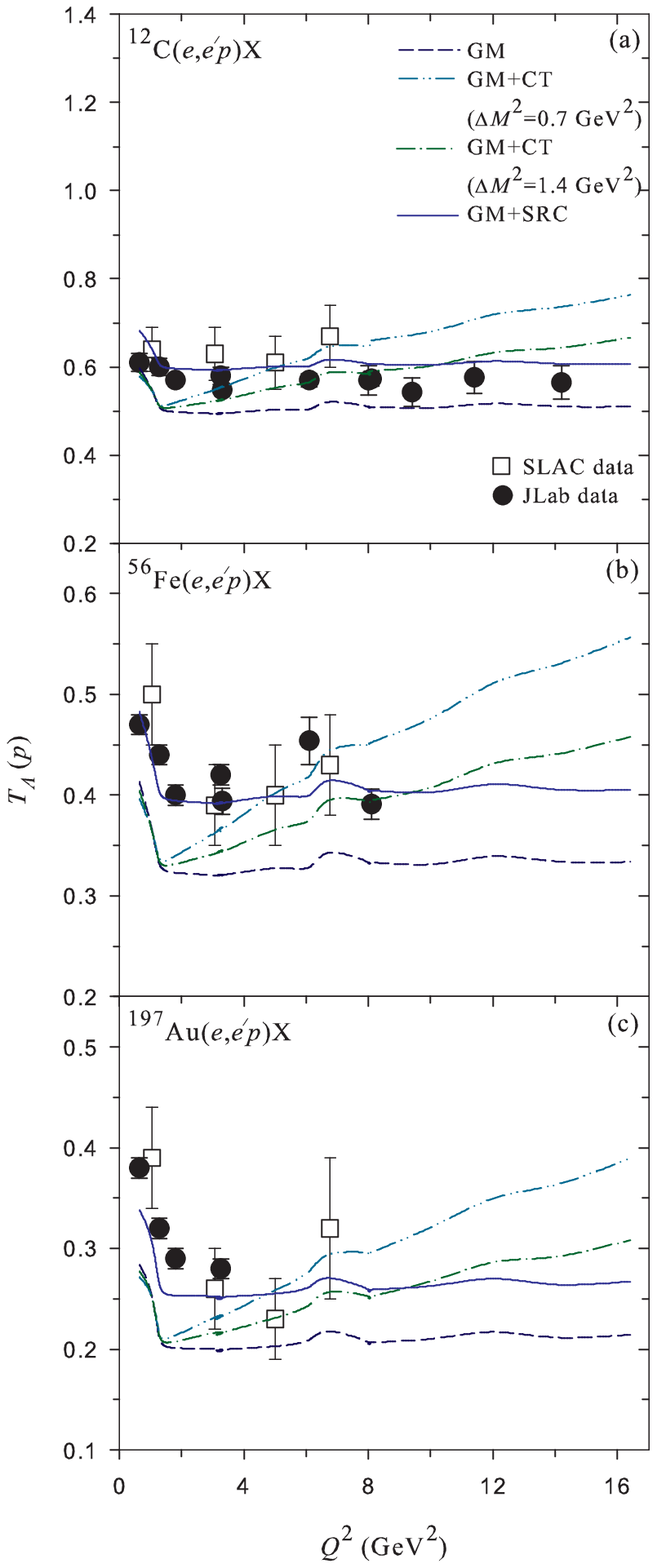,height=14.0 cm,width=06.0 cm}
}}
\caption{
(color online).
The calculated nuclear transparency of the proton $T_A(p)$ vs. photon
virtuality $Q^2$. The dashed curve denotes $T_A(p)$ calculated using Glauber
model (GM).
The
dot-dot-dashed and dot-dashed curves illustrate the proton color transparency
($p$CT) for two different values of $\Delta M^2$, see text. The solid curves
arise due to the inclusion of the short-range correlation (SRC) in GM. The
data are taken from Refs.~\cite{Neill}-\cite{Bhetuwal}.
}
\label{TQpp}
\end{center}
\end{figure}

The calculated proton transparency $T_A(p)$ vs photon virtuality $Q^2$ in
the A$(e,e^\prime p)$X reaction is compared with the data in Fig.~\ref{TQpp}.
The data reported from SLAC \cite{Neill} and JLab \cite{DuttaG, Bhetuwal}
are represented by the white squares and black circles respectively.
Fig.~\ref{TQpp}(a)
shows CT does not exist for the proton moving through $^{12}$C for a wide
range of the photon virtuality, i.e., $0.64 \le Q^2 \le 14.2$ GeV$^2$.
This
is corroborated by the results for other nuclei shown in Figs.~\ref{TQpp}(b)
and (c), where the data are available for lesser range of $Q^2$, i.e.,
$0.64 \le Q^2 \le 8.1$ GeV$^2$ for $^{56}$Fe and $0.64 \le Q^2 \le 6.77$
GeV$^2$ for $^{197}$Au. Therefore, the CT of proton is distinctly ruled out.
Fig.~\ref{TQpp}
shows the calculated $T_A(p)$ due to the inclusion of SRC in GM reproduce
the data reasonably well for all nuclei.

\begin{figure}[h]
%\begin{figure}
\begin{center}
\centerline {\vbox {
%\psdraft
\psfig{figure=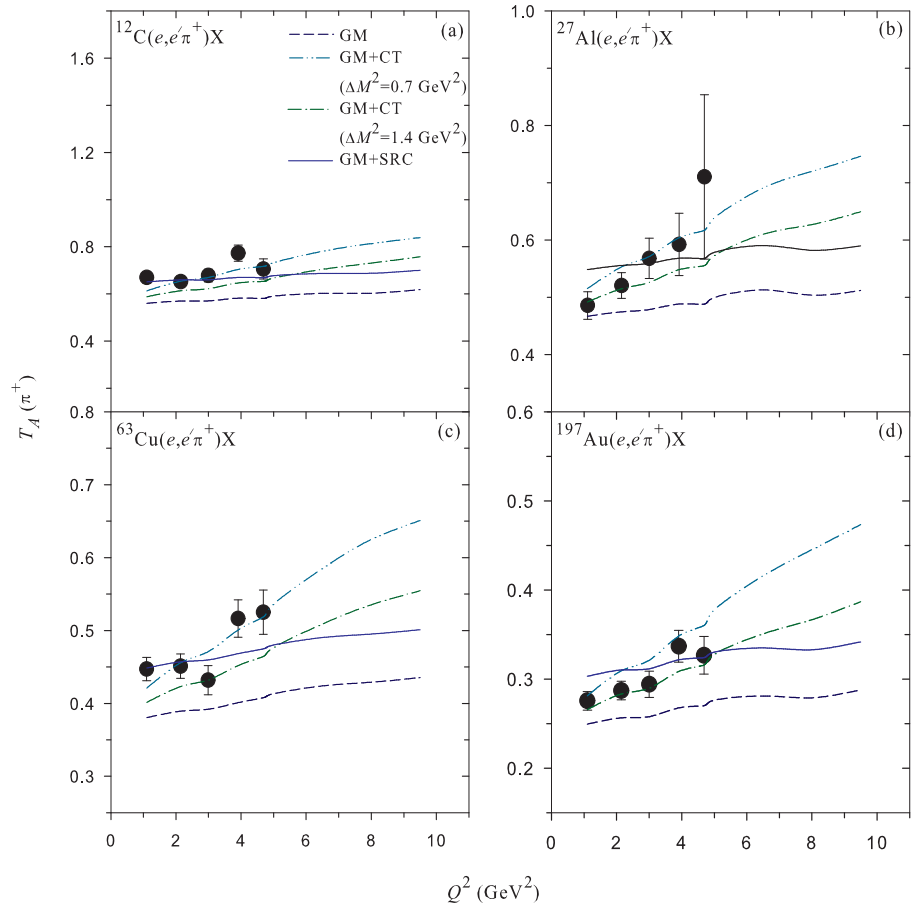,height=12.0 cm,width=12.0 cm}
}}
\caption{
(color online).
Same as those presented in Fig.~\ref{TQpp} but for the pion. The data are
taken from Refs.~\cite{Clasie}.
}
\label{TQppi}
\end{center}
\end{figure}

The measured pionic transparency $T_A(\pi^+)$ for $1.1 \le Q^2 \le 4.69$
GeV$^2$ in the A$(e,e^\prime \pi^+)$X reaction have been reported from JLab
\cite{Clasie} for $^{12}$C, $^{27}$Al, $^{63}$Cu and $^{197}$Au nuclei.
The data for all nuclei (except $^{12}$C) show the enhancement of
$T_A(\pi^+)$ with $Q^2$. Proposals are there to measure $T_A(\pi^+)$ at JLab
for higher $Q^2$, i.e., $5 \le Q^2 \le 9.5$ GeV$^2$ \cite{DuttaH, DuttaP}.
Therefore,
$T_A(\pi^+)$ for $1.1 \le Q^2 \le 9.5$ GeV$^2$ have been calculated and those
are presented in Fig.~\ref{TQppi} along with the available data \cite{Clasie}.
The
calculated results due to GM+CT (i.e., $\pi$CT) are accord with both the
$Q^2$ dependence and magnitude of the data. This is in concurrence with the
earlier calculations \cite{Larionov, Kaskulov}.
The
calculated $T_A(\pi^+)$ due to GM+SRC do not describe the $Q^2$ dependence
of the data but the calculated results agree with the large number of data
points within the errors.
Therefore,
the data of $T_A(\pi^+)$ in the region of $Q^2 = 5-9.5$ GeV$^2$ are necessary
to prove the existence of $\pi$CT.

%\section{Conclusions}

%The nuclear transparencies $T_A$ of the proton and $\pi^+$ meson produced in
%the inclusive $(e,e^\prime)$ reaction on nuclei have been calculated using
%Glauber model for the wide range of photon virtuality $Q^2$. $T_A$ evaluated
%using Glauber model do not reproduce the measured transparencies for both
%proton and pion.
%To
%realize the data, $T_A$ is calculated incorporating the color transparency
%of the produced hadron and the short-range correlation of the bound nucleon
%in the Glauber model.
%The
%calculated transparencies of proton $T_A(p)$ compared with the data show the
%color transparency does not exist for the proton. The calculated $T_A(p)$
%due to the short-range correlation added in Glauber model reproduce the data.
%The
%calculated transparencies of pion $T_A(\pi^+)$ considering the color
%transparency in the Glauber model agree with both the $Q^2$ dependence and
%magnitude of the data, available for $Q^2 = 1.1-4.69$ GeV$^2$.
%The
%calculated $T_A(\pi^+)$ due to the inclusion of the short-range correlation
%in Glauber model agree with the large number of data within the errors, but
%those do not explain the $Q^2$ dependence of the measured spectra.
%Therefore,
%$T_A(\pi^+)$ for $5 \le Q^2 \le 9.5$ GeV$^2$ to be measured at JLab are
%required to confirm the pionic color transparency.

%\section{Acknowledgement}

The author appreciates Prof. Dipnagkar Dutta for the discussions on the
experimental results, and thanks A. K. Gupta and S. M. Yusuf for their
encouragement to work on theoretical nuclear physics.

%\newpage

%\newpage

%{\bf Figure Captions}
%\begin{enumerate}

%\item
%(color online).

%\item
%(color online).

%\end{enumerate}

\newpage
%\vspace{1 cm}

\newpage
%\vspace{1 cm}

\end{document}